\documentclass[a4paper,11pt]{article}
\usepackage{amssymb,amsmath,bm}  
\usepackage{graphics,graphicx}   
\usepackage{array,booktabs}      
\usepackage{authblk}  

\usepackage{cite}
\usepackage{xcolor}
\usepackage[margin=2.5cm]{geometry}

\usepackage[linktocpage,
colorlinks = true,
linkcolor = blue,
urlcolor  = blue,
citecolor = red,
anchorcolor = blue]{hyperref}

\usepackage{epsfig}

\numberwithin{equation}{section}

\title{\textbf{W-boson mass anomaly and vacuum structure in vector dark matter model with a singlet scalar mediator}}
\author[a]{Seyed Yaser Ayazi\thanks{syaser.ayazi@semnan.ac.ir}}
\author[a]{ Mojtaba Hosseini\thanks{Mojtabahosseini743@yahoo.com}}
\affil[a]{Physics Department, Semnan University, P.O. Box. 35131-19111, Semnan, Iran}
\date{\today}

\begin{document}

\baselineskip 0.6 cm

\maketitle

\begin{abstract}

Motivated by the deviation of the W boson mass reported by the CDF collaboration, we study an extension of the Standard Model (SM) including a vector dark matter (VDM) candidate and a scalar mediator. In the model, the one-loop corrections induced by the new scalar, shift the W boson mass. We identify the parameter space of the model consistent with dark matter (DM) relic abundance, W mass boson anomaly, invisible Higgs decay at LHC, and direct detection of DM. It is shown that the W-mass anomaly can be explained for the large part of parameter space of VDM mass and scalar mediator mass between $100-124~\rm GeV$ by the model. We also investigate the renormalization group equations (RGE) at one-loop order for the model. We show that the contribution of new scalar mediator to RGE,   guarantees positivity and vacuum stability of SM Higgs up to Planck scale.

\end{abstract}

\section{Introduction} \label{sec1}
 Recently, CDF-II collaboration published a new result in  W boson mass with increased precision  $M^W_{CDF} =80.4335 \pm 0.0094 ~\rm{GeV}$ which deviates from SM prediction by $7\sigma$\cite{CDF:2022hxs}. The SM prediction for the W boson mass is $ M^W_{SM}=80.357 \pm 0.006 ~\rm {GeV}$ \cite{ParticleDataGroup:2020ssz}.
Needless to say a better understanding of SM calculations, and also more accurate  measurements are needed. Nevertheless, a lot of new suggestions have been proposed to explain  this anomaly \cite{Ghorbani:2022vtv,Cheng:2022aau,Borah:2022zim,Arcadi:2022dmt,Nagao:2022oin,Kawamura:2022uft,Fan:2022dck,Lu:2022bgw,Athron:2022qpo,Yuan:2022cpw,Strumia:2022qkt,Yang:2022gvz,deBlas:2022hdk,Du:2022pbp,Tang:2022pxh,Cacciapaglia:2022xih,Blennow:2022yfm,Sakurai:2022hwh,Fan:2022yly,Liu:2022jdq,Lee:2022nqz,Bagnaschi:2022whn,Paul:2022dds,Bahl:2022xzi,Asadi:2022xiy,DiLuzio:2022xns,Athron:2022isz,Gu:2022htv,Babu:2022pdn,Heo:2022dey,Du:2022brr,Cheung:2022zsb,Crivellin:2022fdf,Endo:2022kiw,Biekotter:2022abc,Balkin:2022glu,Han:2022juu,Ahn:2022xeq,Zheng:2022irz,Ghoshal:2022vzo,FileviezPerez:2022lxp,Mondal:2022xdy,Borah:2022obi,Chowdhury:2022moc,Bhaskar:2022vgk,Lazarides:2022spe,Senjanovic:2022zwy,Chowdhury:2022dps,Heckman:2022the,Cao:2022mif,Zhou:2022cql}.

In this paper, we propose a possible explanation of the W boson mass anomaly as well as the nature of dark matter with a extra $U(1)$ dark sector.
In\cite{Zhang:2022nnh}, the extra $U(1)$ gauge field mix with the $U(1$) hypercharge via gauge kinetic term,  and this kinetic mixing can generate an enhancement of the W boson mass. Another approach for the $U(1)$ extension of SM is to consider an additional scalar field in which the new field can enhance W-boson mass via the loop corrections\cite{Duch:2015jta}. In the model, the vector dark field does not mix with the SM gauge field and a scalar field plays the role of mediator between the dark sector and SM. This field shift W-boson mass in loop corrections. We study this simple extension of the SM to explain the W boson mass enhancement and also offer a viable DM candidate with mass ranging from $1~\rm GeV$ to $2~\rm TeV$. In the following, we apply various phenomenological constraints such as invisible Higgs decay mode and direct detection experiment in our analysis.

In the context of SM, the vacuum stability and perturbativity have resulted in theoretical bounds on the Higgs mass. The Higgs mass value, together with other relevant parameters such as the top quark mass, affects on the behavior of  Higgs potential at very high energy scales, in particular for the sake of electroweak vacuum stability \cite{Bezrukov:2012sa},\cite{Buttazzo:2013uya}.
This is because, for Higgs and top mass values, the Higgs quartic coupling can be very small or even negative. Since the top Yukawa coupling dependency is strong and very subtle, there are different views on an explanation of this issue in literature, some of them favoring\cite{Bezrukov:2012sa} and some others disfavoring \cite{Degrassi:2012ry}. In the following, we also ask how the model behaves at a high energy scale and is it computationally reliable?  Standard treatment is the study of the running of the coupling constants in terms of the mass scale $\Lambda$ via the RGE. The three standard problems to consider are the positivity, perturbativity of the coupling constants and the vacuum stability  of the model. These issues have been studied in the literature in the presence of a scalar extension of SM \cite{Ghorbani:2017qwf,Gonderinger:2012rd} and it was shown that the vacuum stability requirement can affect the DM relic density. Here, we discuss the requirement of vacuum stability and RGE of the model.

This work is organized as follows: After the introduction, we introduce the model.
In section~3, we study conditions of vacuum stability and calculate the RGEs of the model. In section~4,  we study the contribution of the model to W-boson mass and probe the consistent parameter space of the model with CDF-II measurement. Invisible Higgs decay constraint on the model study in section.~5. In section.~6, we find the allowed regions in the parameter space which will give rise to the correct DM relic density. We describe combined results  consistent with constraints provided and examine the RGEs numerically in section~7 . Section 8 contains our conclusions.

\section{The Model} \label{sec2}
In our model, beyond the SM, we employ two new fields to furnish the model: a  complex scalar field $S$ which has a unit charge under a dark $U(1)$ gauge symmetry with a dark photon vector field $ V_{\mu} $. The model has an additional $Z_2$ discrete symmetry, under which the vector field $ V_{\mu} $ and the scalar
field transform as follows: $V_{\mu} \rightarrow - V_{\mu}$, $S\rightarrow S^*$ and all the other fields are even.  $Z_2$ symmetry forbids the kinetic mixing between the vector field $ V_{\mu} $ and SM $ U_{Y}(1) $ gauge boson $ B_{\mu} $, i.e., $ V_{\mu \nu} B_{\mu \nu} $. Therefore, the vector field $ V_{\mu} $ is stable and can be considered a DM candidate. The Lagrangian one can write with assumption is:
\begin{equation}
 {\cal L} ={\cal L}_{SM} + (D'_{\mu} S)^{*} (D'^{\mu} S) - V(H,S) - \frac{1}{4} V_{\mu \nu} V^{\mu \nu} , \label{2-2}
\end{equation}
where $ {\cal L} _{SM} $ is the SM Lagrangian without the Higgs potential term and
\begin{align}
& D'_{\mu} S= (\partial_{\mu} + i g_v V_{\mu}) S,\nonumber \\
& V_{\mu \nu}= \partial_{\mu} V_{\nu} - \partial_{\nu} V_{\mu},\nonumber \end{align}
and the potential which is renormalizable and invariant
under gauge and $ Z_{2} $ symmetry is:
\begin{equation}
V(H,S) = -\mu_{H}^2 H^{\dagger}H-\mu_{S}^2 S^*S+\lambda_{H} (H^{\dagger}H)^{2} + \lambda_{S} (S^*S)^{2} +  \lambda_{S H} (S^*S) (H^{\dagger}H). \label{2-3}
\end{equation}
Note that the quartic portal interaction, $ \lambda_{SH} (S^*S) (H^{\dagger}H) $, is the only connection between the dark sector and the SM.

SM Higgs field $ H $, as well as dark scalar $S$, can receive VEVs breaking respectively the electroweak and $ U'_{D}(1) $ symmetries.
In the unitary gauge, the imaginary component of $S$ can be absorbed as the longitudinal component of $ V_{\mu} $.
In this gauge, we can write
\begin{equation}
H = \frac{1}{\sqrt{2}} \begin{pmatrix}
0 \\ h_{1} \end{pmatrix} \, \, \, {\rm and} \, \, \, S = \frac{1}{\sqrt{2}} h_{2} , \label{2-4}
\end{equation}
where $ h_{1} $ and $ h_{2} $ are real scalar fields which can get VEVs.
The tree level potential in unitary gauge is:
\begin{equation}
V_{\text{tree}}(h_{1},h_{2})=-\frac{1}{2} \mu _H^2 h_1^2-\frac{1}{2} \mu _S^2 h_2^2 +\frac{1}{4} \lambda _H h_1^4 +\frac{1}{4} \lambda _S h_2^4+\frac{1}{4} \lambda _{SH} h_1^2 h_2^2.
\end{equation}
Given differentiable $ V_{\text{tree}} $, one can obtain
the Hessian matrix, $ {\cal{H}}_{ij}(h_{1},h_{2})=\frac{\partial^{2}V_{\text{tree}}}{\partial h_{i} \partial h_{j} }$.
In order to get the mass spectrum of the model, it is necessary to consider the sufficient conditions for a local minimum:
\begin{align}
& \quad \dfrac{\partial V_{\text{tree}}}{\partial h_1}\bigg|_{h_1=h_2=0}=0 ,\\
& \quad \dfrac{\partial V_{\text{tree}}}{\partial h_2}\bigg|_{h_1=h_2=0}=0 \label{minimum1} ,\\
&  \det {\cal{H}} > 0 \label{minimum2} \\
& {\cal{H}}_{11} > 0 \label{minimum3}
\end{align}
to occur at a point $ (\nu_{1},\nu_{2}) $. Note that Eq. (\ref{minimum2}) and Eq. (\ref{minimum3}) also imply that $ {\cal{H}}_{22} > 0 $.
Eq. (\ref{minimum1}) leads to
\begin{align}
& \mu _H^2= \lambda _H \nu _1^2+ \frac{1}{2} \lambda _{SH} \nu _2^2 , \nonumber\\
& \mu _S^2=\lambda _S \nu _2^2+ \frac{1}{2} \lambda _{SH} \nu _1^2  \label{mini}
\end{align}
Eq. (\ref{mini}) leads to the non-diagonal mass matrix $\cal{H}$ as follows:
\begin{equation}
{\cal{H}}(\nu_{1},\nu_{2})= \left(
\begin{array}{cc}
 2 \lambda _H \nu _1^2 & \lambda _{SH} \nu _1 \nu _2 \\
 \lambda _{SH} \nu _1 \nu _2 & 2 \lambda _S \nu _2^2 \\
\end{array}
\right) \label{hess}
\end{equation}
Therefore, according to the conditions $ {\cal{H}}_{11} > 0 $ and $ {\cal{H}}_{22} > 0 $ and Eq. (\ref{minimum2}) we should have
\begin{equation}
\lambda_{H} > 0 \, \, \, , \, \, \,  \lambda_{S} > 0 \, \, \, , \, \, \, \lambda_{SH}^{2} < 4 \lambda_{H} \lambda_{S}
\end{equation}
Now by substituting $ h_1 \rightarrow \nu_1 + h_1 $ and $ h_2 \rightarrow \nu_2 + h_2 $, the fields $ h_1 $ and $ h_1 $ mix with each other and they can be rewritten by the mass eigenstates $ H_1 $ and $ H_1 $ as
\begin{equation}
\begin{pmatrix}
h_{1}\\h_{2}\end{pmatrix}
 =\begin{pmatrix} cos \alpha~~~  sin \alpha \\-sin \alpha  ~~~~~cos \alpha
 \end{pmatrix}\begin{pmatrix}
H_1 \\  H_{2}
\end{pmatrix}, \label{matri}
\end{equation}
where $ \alpha $ is the mixing angle. After symmetry breaking, we have
\begin{align}
& \nu _2=\frac{M_V}{g_v} \nonumber,~~~~~~~~~~ \sin\alpha=\frac{\nu_1}{\sqrt{\nu _1^2+\nu_2^2}} \\
& \lambda _H=\frac{\cos ^2\alpha M_{H_1}^2+\sin ^2\alpha  M_{H_2}^2}{2 \nu _1^2}  \nonumber \\
& \lambda _S=\frac{\sin ^2\alpha M_{H_1}^2+\cos ^2\alpha  M_{H_2}^2}{2 \nu _2^2}  \nonumber \\
&  \lambda _{SH}=\frac{ \left(M_{H_2}^2-M_{H_1}^2\right) \sin \alpha  \cos \alpha}{\nu _1 \nu _2} \label{cons}
\end{align}

Since $U(1)$ gauge symmetries of two sectors are only broken spontaneously, gauge bosons from the two sectors will not mix at any order of perturbation theory and the field renormalisation constants are defined independently in each sector \cite{Glaus:2019itb}. This means vector dark matter still is stable and DM candidate.

The mass eigenstates of scalar fields can be written as following:
\begin{align}
M^2_{H_{2},H_{1}}=\lambda_H \nu_1^2+\lambda_S \nu_2^2 \pm \sqrt{(\lambda_H \nu_1^2-\lambda_S \nu_2^2)^2+\lambda_{SH}^2\nu_1^2\nu_2^2},
\label{mas}
\end{align}
where we take $ M_{H_1} = 125 $ GeV and $ \nu _1 = 246 $ GeV. Note that, beside of SM parameters, the model has only three free parameters $ g_v $, $ M_{H_2} $ and $ M_V $.

\section{Vacuum stability and RGE}

A prominent feature of the study of high energy physics is the evolution of the coupling constants with energy. This has become an incentive to further strengthen theories such as the GUT and supersymmetry by merging couplings at high energies\cite{Wulzer:2019max}. Renormalization Group Equation(RGE) describes the behavior of quantities with energy. After the discovery of the Higgs particle by ATLAS and CMS experiments at the LHC in 2012, the vacuum stability study has been done more clearly\cite{Ghorbani:2017qwf,Gonderinger:2012rd,Abada:2013pca,Baek:2012uj,Ghorbani:2021rgs,Duch:2015jta}. In the SM, the Higgs quartic coupling becomes negative at the scale $10^{10}$ GeV, and the Higgs non-zero VEV is no longer a minimum of the theory. The reason for this is that the top quark has a large negative contribution to the RGE for $\lambda_{H}$.  We  will show the running of $\lambda_{H}$ in SM in the next sections.

There are three types of theoretical constraints on the couplings in the model.  The first is the perturbative unitarity condition of the couplings, which includes the following relations\cite{Hashino:2018zsi}:
\begin{equation}
| \lambda_{S}|<4\pi \, \, \, , \, \, \, | \lambda_{H}|<4\pi \, \, \, , \, \, \, | \lambda_{SH}|<8\pi \, \, \, , \, \, \, 3\lambda_{H}+2\lambda_{S}+\sqrt{(3\lambda_{H}-2\lambda_{S})^2 +2\lambda_{SH}^2 } <8\pi \, \, \, , \, \, \,| g_v|<\sqrt{4\pi}
\end{equation}
 and the second is vacuum stability of the model dictates some other constraints on the couplings such that for self-coupling constants. In this regard, we should have $\lambda_i$ and $\lambda_H>0$. On the other hand, by adding a new scalar mediator the vacuum structure of the model will be modified. The third condition is positivity where potential must be well-defined and positive at all scales. The requirement of positivity for the potential implies the following relations :
\begin{equation}
\lambda_H>0   ,  \lambda_S>0  , \lambda_{SH}>-2 \sqrt{\lambda_H \lambda_S }
\end{equation}
Also, the common investigation for vacuum stability analyses in the literature begins with the RGE improved  potential and choice of the renormalization scale to minimize  the one-loop potential. In this light, we  consider the running of the coupling constants with energy. The Model is implemented in SARAH \cite{Staub:2015kfa} to compute $\beta$ functions and their runnings. We calculate the one-loop RGE and one-loop $\beta$ functions for scalar couplings and dark coupling including the following relationships:
\begin{align}
& (16\pi^2)\beta_{\lambda_{S}} = -20\lambda_{S}^2 -2\lambda_{SH}^2 -6g_v^4 -12g_v^2 \lambda_{S} ,\nonumber \\
& (16\pi^2)\beta_{\lambda_{SH}}= -\frac{3}{2}g_{1}^2 \lambda_{SH} -\frac{9}{2}g_{2}^2 \lambda_{SH} -12\lambda_{SH}\lambda_{H} -8\lambda_{SH}\lambda_{S} -4\lambda_{SH}^2 + 6\lambda_{SH}\lambda_{t}^2 -6g_v^2 \lambda_{SH}  \nonumber ,\\
& (16\pi^2)\beta_{\lambda_{H}}= -\frac{3}{8}g_{1}^4 -\frac{3}{4}g_{1}^2 g_{2}^2 -\frac{9}{8}g_{2}^4 -3g_{1}^2 \lambda_{H} -9g_{2}^2 \lambda_{H} -24\lambda_{H}^2 -\lambda_{SH}^2 +12\lambda_{H}\lambda_{t}^2 +6\lambda_{t}^4, \nonumber \\
& (16\pi^2)\beta_{g_v}= \frac{1}{3}g_v^3.
\label{RGE}
\end{align}
where $\lambda_t$ is top Yukawa coupling and $\beta_a\equiv \mu \frac{da}{d\mu}$ that $\mu$ is the renormalization scale with initial value $\mu_0=100~\rm GeV$. The $\beta$ functions of couplings, $g_1$, $g_2$ and $g_3$ are given to one-loop order by:
\begin{align}
& (16\pi^2)\beta_{g_{1}}= \frac{41}{6} g_{1}^3 , \nonumber \\
& (16\pi^2)\beta_{g_{2}}= -\frac{19}{6} g_{2}^3 , \nonumber \\
& (16\pi^2)\beta_{g_{3}}= -7 g_{3}^3 .
\end{align}
Among the Yukawa couplings of SM, the top quark has the largest contribution compared with other fermions in the SM. Therefore, we set all the SM Yukawa couplings equal to zero and consider only the top quark coupling. The RGE of top quark Yukawa coupling is given to one-loop order by
\begin{align}
(16\pi^2)\beta_{\lambda_t}= -\frac{17}{12}g_{1}^2 \lambda_t -\frac{9}{4}g_2^2 \lambda_{t} -8g_3^2 \lambda_t +\frac{9}{2}\lambda_t^3 .
\end{align}

In the following, we first study experimental constraints on the model, and then in the final section, we investigate conditions of the Higgs stability and RGE of coupling parameters of the model.

\section{$\rm W$-Mass Anomaly}
To explain the CDF-II anomaly, we study W-mass correction in the context of the model. The corrections of new physics to the W-boson mass can be written in terms of the Peskin-Takeuchi oblique parameters $S, T,$ and $U$. The Peskin–Takeuchi parameters are only sensitive to new physics that contribute to the oblique corrections, i.e., the vacuum polarization corrections to four-fermion scattering processes. In general, the SM contribution to an oblique parameter  is subtracted from the new physics contribution  to define the oblique parameter. The effects of $S, T,$ and $U$ on $W$ boson mass can be expressed as follows
 \cite{Peskin:1991sw}:
\begin{equation}
\Delta M_W^2 = \frac{ M_{SM}^2}{c_W^2 - s_W^2} \left(
-\frac{ \alpha S}{2}  + c_W^2  \alpha T +\frac{c_W^2-s_W^2}{4s_W^2}   \alpha U \right)\,.
\label{WMco}
\end{equation}
where $M_{SM}$ is SM of W-boson mass, $c_W$ and $s_W$ are cosine and sine of Weinberg angle. Oblique parameters $S, T,$ and $U$ for our model are as follows \cite{Grimus:2008nb} :
\begin{equation}
\alpha S=\frac{g^2 sin ^2\alpha}{96 {\pi}^2 }[(ln {M_{H_2}}^2 + G( {M_{H_2}}^2 ,{M_{Z}}^2 ))-(ln {M_{H_1}}^2 + G({M_{H_1}}^2 ,{M_{Z}}^2 )) ] ,
\end{equation}
\begin{equation}
\alpha T=\frac{3g^2 sin ^2\alpha}{64\pi^2 s_W^2 M_W^2 }[(F(M_Z^2 ,M^2_{H_2}) - F(M_W^2 ,M_{H_2}^2)) - (F(M_Z^2 ,M_{H1}^2)-F(M_W^2 ,M_{H_1}^2))] ,
\end{equation}
\begin{equation}
\alpha U=\frac{g^2  sin ^2\alpha}{96\pi^2 }[(G( M_{H_2}^2 ,M_W^2 )-G( M_{H_2}^2 ,M_{Z}^2 ))-(G(M_{H_1}^2 ,M_{W}^2 )-G(M_{H_1}^2 ,M_{Z}^2))] ,
\end{equation}
where
\begin{equation}
g^2= 4\pi \alpha_{QED}
\end{equation}
\begin{eqnarray}
F(x,y)=\bigg\lbrace \begin{array}{cc}
{\frac{x+y}{2}-\frac{xy}{x-y}ln\frac{x}{y}}&{for~x\neq y ,}\\{0}&{for~ x=y ,}
\end{array} ,
\end{eqnarray}
\begin{equation}
G(x,y)=-\frac{79}{3}+ 9\frac{x}{y} -2\frac{x^2}{y^2} +(-10+18\frac{x}{y}-6\frac{x^2}{y^2}+\frac{x^3}{y^3}-9\frac{x+y}{x-y}) ln\frac{x}{y} +(12-4\frac{x}{y}+\frac{x^2}{y^2}) \frac{f(x,x^2 -4xy)}{y} ,
\end{equation}
\begin{eqnarray}
f(a,b)=\bigg\lbrace \begin{array}{cc}
{\sqrt b ln\vert \frac{a-\sqrt b}{a+\sqrt b}\vert}&{for~ b>0}\\{0}&{for~ b=0}\\{2\sqrt{-b}~ Arctan \frac{\sqrt{-b}}{a}}&{for~ b<0} .
\end{array} ,
\end{eqnarray}

\begin{figure}
	\begin{center}
		\centerline{\hspace{0cm}\epsfig{figure=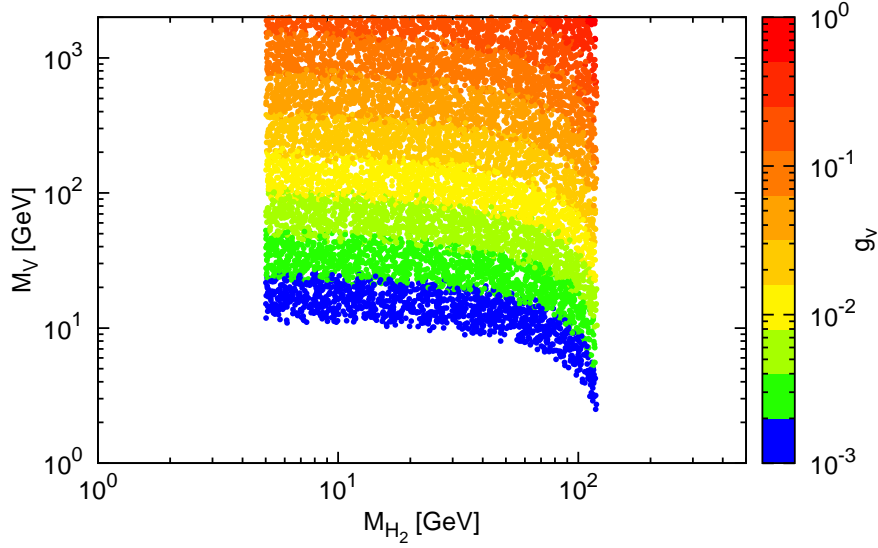,width=12cm}}
		\centerline{\vspace{-0.2cm}}
		\caption{Scatter points depict allowed range of parameters space of the model consistent with W-boson mass measurement.} \label{Wmass}
	\end{center}
\end{figure}

 We know that in the model $U(1)$ dark gauge symmetry have been broken spontaneously and $Z_2$ still is symmetry of
the model and so there is no mixing between SM and dark gauge bosons\cite{Glaus:2019itb}. Meanwhile, as it is clear from the above relations extra gauged $U(1)$ symmetry does not contribute to W- mass anomaly and contributions are the same as those found
in the analysis of the plain Higgs portal\cite{Duch:2015jta}. but, contributions of parameters $M_V$ and $g_v$ appear
in the form of $sin \alpha$ in the formulas of (4.2), (4.3) and (4.4). To study the model parameter space, other than theoretical constraints (such as perturbativity condition and vacuum stability), it is necessary to consider the experimental upper limit on mixing angle $\alpha$. For low masses, $M_{H_2} < 5~\rm GeV$, the strongest limit comes from  decay $B\rightarrow K\ell\ell$\cite{LHCb:2012juf,Belle:2009zue}. It was shown that for this range of parameters, $\sin\alpha$ should be smaller than $10^{-3}$. Between $5-12~\rm GeV$, the constraint on $sin \alpha< 0.5$ is imposed by the decay of a low-mass Higgs boson in radiative decay of the Y and the DELPHI searches for a light Higgs in Z-decay\cite{DELPHI:1990vtb,BaBar:2012wey}. For above this mass range  to about $65~\rm GeV$, an overall result of the Higgs signal strength measured by ATLAS and CMS \cite{ATLAS:2016neq} severely constrains the mixing angle  to values smaller than $sin\alpha<0.12$. For the larger values of $M_{H_2}$, the lower limit is set by the LHC constraints on the mixing angle in which $sin \alpha\leq0.44$\cite{Farzinnia:2013pga,Farzinnia:2014xia}. In the following, for low mass $M_{H_2}$ in which decay of SM Higgs like to $H_2$ is kinematically  possible, we consider severe upper limit on mixing angle and for large $M_{H_2}$ mass, we relax this bound. As was discussed in the previous section, the present model has three free parameters, $ g_v $, $ M_{H_2} $ and $ M_V$.  In addition to the mixing angle constraints, we also make the following choices for the mass parameters:
 \begin{itemize}
 	\item The DM mass $M_V$ is between $1-2000~\rm GeV$;
 	\item The mediator scalar mass ($M_{H_2}$) is between $1-500~\rm GeV$;
 \end{itemize}
We scan over the three-dimensional parameters $ g_v $, $ M_{H_2} $, and $ M_V $ to probe a consistent range of parameters space with observables.
  In figure.~\ref{Wmass}, we depict the allowed range of parameters of the model which is consistent with CDF measurement for W-boson mass. Note that, for a large value of $M_{H_2}$, we choose $sin\alpha\leq0.44$ on the mixing angle, and for $M_{H_2}<65~\rm GeV$, we suppose $\alpha<6.9^{~\circ}$ to satisfy ATLAS and CMS upper limit on the mixing angle. As is seen in the figure, the $\rm W$-Mass measurement, for $M_{H_2}\lesssim 4.5~\rm GeV$ and $M_{H_2}\gtrsim 124~\rm GeV$, excludes  the parameters space of the model. However, for $M_{H_2}$ between $4.5-124~\rm GeV$ and $M_V$  between $1-2000~\rm GeV$, the model is consistent with the W-mass anomaly.

\section{Invisible Higgs decay}
In the model, SM Higgs-like can decay invisibly into a pair of DM if kinematically allowed. Also, it can decay to another Higgs boson for $M_{H_2}<1/2M_{H_1}$. Therefore, $H_1$ can contribute to the invisible decay mode with a branching ratio:
\begin{eqnarray}
Br(H_1\rightarrow \rm Invisible)& =\frac{\Gamma(H_1\rightarrow 2VDM)+\Gamma(H_1\rightarrow 2H_2)}{\Gamma(h)_{SM}+\Gamma(H_1\rightarrow 2VDM)+\Gamma(H_1\rightarrow 2H_2)},
\label{decayinv1}
\end{eqnarray}

where $\Gamma(h)_{SM}=4.15 ~ \rm [MeV]$ is total width of Higgs boson \cite{LHCHiggsCrossSectionWorkingGroup:2011wcg}. The partial width for processes $H_1\rightarrow 2VDM$ and $H_1\rightarrow 2H_2$ are given by:
\begin{eqnarray}
\Gamma(H_1\rightarrow 2VDM)& =\frac{g_v^4v^2_2 sin^2{\alpha}}{8\pi M_{H_1}}\sqrt{1-\frac{4M^2_{V}}{M^2_{H_1}}}.
\label{decayinv1}
\end{eqnarray}

\begin{eqnarray}
\Gamma(H_1\rightarrow 2H_2)& =\frac{a^2}{8\pi M_{H_1}}\sqrt{1-\frac{4M^2_{H_2}}{M^2_{H_1}}}.
\label{decayinv1}
\end{eqnarray}

where $a=(1/2cos^3\alpha -sin^2\alpha cos\alpha)v_1$.
The SM prediction for the branching ratio of the Higgs boson decaying to invisible particles which coming from process $h\rightarrow ZZ^*\rightarrow 4\nu$ \cite{Denner:2011mq},\cite{Dittmaier:2012vm},\cite{Brein:2003wg},\cite{LHCHiggsCrossSectionWorkingGroup:2013rie} is, $1.2\times10^{-3}.$
CMS Collaboration has  reported the observed (expected) upper limit
on the invisible branching fraction of the Higgs boson to be $0.18 (0.10)$ at the $95\%$ confidence level, by assuming the SM production cross section \cite{CMS:2022qva}. A Similar analysis was performed by ATLAS collaboration in which an observed upper limit of $0.145$ is placed on the branching fraction of its decay into invisible particles at a $95\%$ confidence level\cite{ATLAS:2022yvh}.

\begin{figure}
	\begin{center}
		\centerline{\hspace{0cm}\epsfig{figure=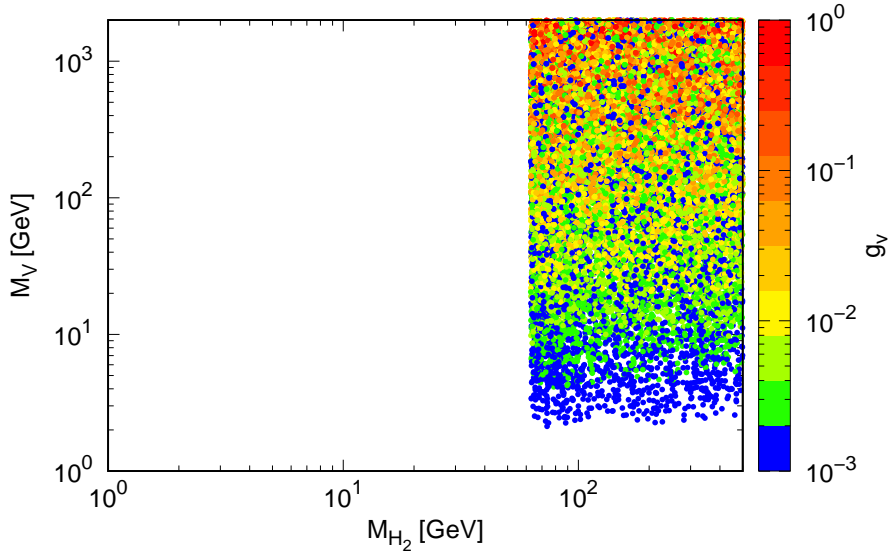,width=12cm}}
		\centerline{\vspace{-0.2cm}}
		\caption{ The cross points depict allowed region which is consistent with invisible Higgs decay at \cite{CMS:2022qva}.
		} \label{Invisible}
	\end{center}
\end{figure}

Figure.~\ref{Invisible}, shows the allowed range of parameters  by considering CMS\cite{CMS:2022qva} upper limit for invisible Higgs mode. In this figure, we consider LHC bound on the mixing angle $\sin\alpha<0.12$. For $M_{H_2}<1/2M_{SM}$  CMS upper limit excludes the parameters space. Note that this upper limit  practically can not constraint the model for low mass DM and also part of the parameters space in which invisible Higgs decay is forbidden.

\section{Relic Density}
\begin{figure}
	\begin{center}
		\centerline{\hspace{0cm}\epsfig{figure=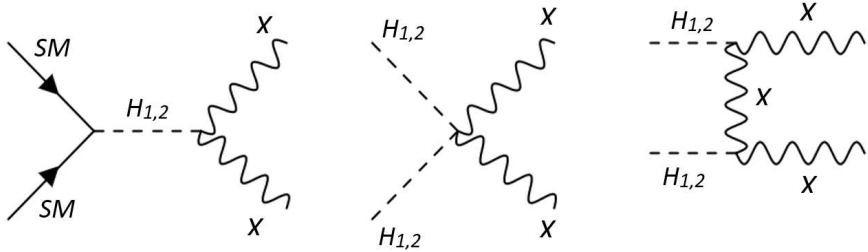,width=12cm}}
		\centerline{\vspace{-0.2cm}}
		\caption{The dominant Feynman diagrams for dark matter relic density production cross section.} \label{feynman}
	\end{center}
\end{figure}

 The evolution of the number density of DM particles $(n_X)$ with time is governed by the Boltzmann equation:
\begin{equation} \label{44}
\dot{n_X} + 3Hn_X = -\langle\sigma_{ann} \nu_{rel}\rangle [n_X ^2 -(n_X ^{eq})^2] ,
\end{equation}
where $H$ is the Hubble parameter and $n_X ^{eq} \sim (m_X T)^{3/2} e^{-m_X /T} $ is the particle density before particles get out of equilibrium. The dominant Feynman diagrams for dark matter production processes are shown in the Fig~.\ref{feynman}.
 In this regard, we calculate the relic density
numerically for the VDM particle by implementing  the model into micrOMEGAs \cite{Belanger:2014vza}. We investigate viable parameter space which satisfies constraints from observed DM relic density ( according to the data of Planck collaboration \cite{Planck:2014egr}):

\begin{equation} \label{44}
\Omega_{DM} h^{2} = 0.1199 \pm 0.0027.
\end{equation}

\begin{figure}
	\begin{center}
		\centerline{\hspace{0cm}\epsfig{figure=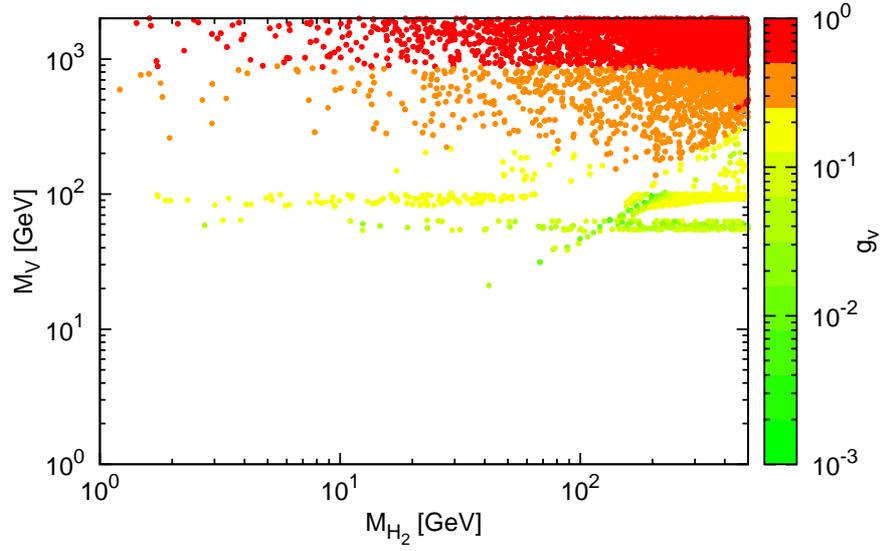,width=12cm}}
		\centerline{\vspace{-0.2cm}}
		\caption{The allowed range of parameter space consistent with DM relic density.} \label{Relic}
	\end{center}
\end{figure}

The allowed range of parameter space corresponding to this constraint is depicted in figure \ref{Relic}. As seen in the figure, for small values of VDM mass ($ M_V\lesssim 22~ \rm GeV$), relic density measurement excludes the model for $10^{-3}\leqslant g_v\leqslant 1$. This issue arises from the fact that we have chosen $g_v$ coupling larger than $10^{-3}$. For smaller values of $g_v$, the model satisfy relic density constraint.

\begin{figure}
	\begin{center}
		\centerline{\hspace{0cm}\epsfig{figure=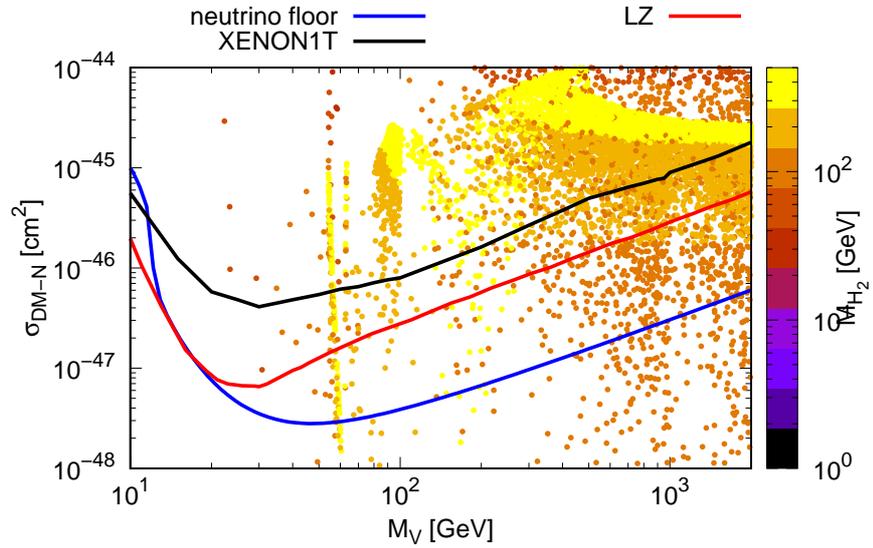,width=12cm}}
		\centerline{\vspace{-0.2cm}}
		\caption{The allowed range of parameter space consistent with DM relic density and DD.} \label{direct detection}
	\end{center}
\end{figure}

\section{Final Results}
Before, we present a combined analysis of all constraints, let us turn our attention to the direct detection(DD) of VDM in the model. At the tree level, a VDM particle can interact elastically with a nucleon either through $H_1$ or via $H_2$ exchange\cite{YaserAyazi:2019caf,YaserAyazi:2018lrv}. Presently, the XENON1T experiment \cite{XENON:2018voc} excludes new parameter space for the WIMP-nucleon spin-independent elastic scatter cross-section above 6 GeV with a minimum of $ 4.1\times10^{-47} cm^{2}$ at 30 GeV.  We restrict the model with these results. Figure \ref{direct detection} shows the parameter space of the model in agreement with the direct detection and relic density constraints. In addition, in this figure we have also used the results of the LUX-ZEPLIN(LZ) experiment \cite{LZ:2022ufs} that was published recently.

\begin{figure}
	\begin{center}
		\centerline{\hspace{0cm}\epsfig{figure=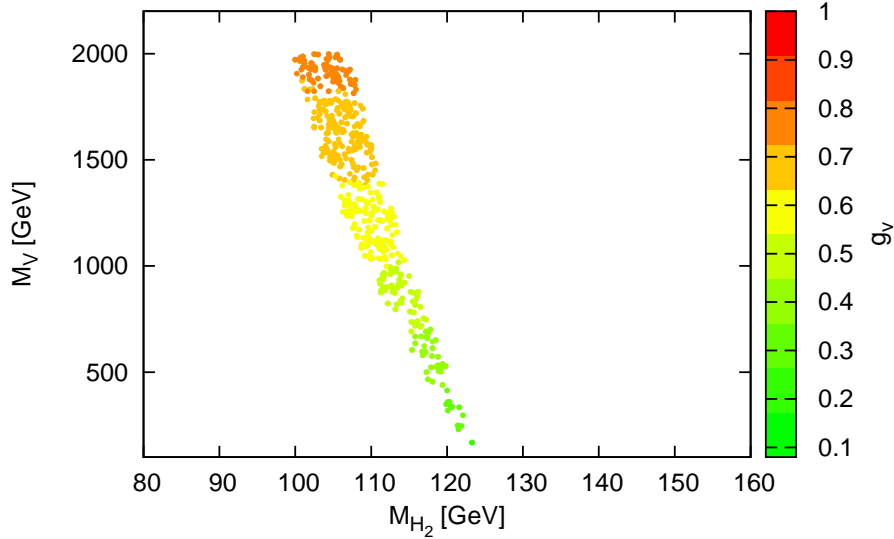,width=12cm}}
		\centerline{\vspace{-0.2cm}}
		\caption{Final results for allowed ranges of parameters of the model. The cross points depict allowed region which is consistent with relic density, direct detection, W mass anomaly and invisible decay rate.} \label{final}
	\end{center}
\end{figure}

The direct detection restrictions and constraints discussed in the previous sections are summarized in figure.~\ref{final}.
The cross points show allowed region consistent with relic density, W-mass anomaly, direct detection as well as the invisible decay rate.

The outcome of imposing these experimental constraints on the model is for a large portion of VDM mass values, narrow region of scalar mediator $H_2$ ($100~\rm GeV \lesssim M_{H_2}\lesssim 124~\rm GeV$), and $0<g_v\leqslant 1$ all observational constraints are satisfied. The point is that  for $M_{H_2}$ and $M_V<1/2M_{SM}$  (in which $H_1$ can decay to $H_2$ and VDM), and $M_{H_2}>124~\rm GeV$ the model is respectively excluded by invisible Higgs upper limit and W-mass CDF-II measurement.

In section.~3, we analyze RGEs of coupling of the model. We show that adding a new VDM and scalar mediator, the RGEs will be modified. It is interesting to see behavior of RGEs of couplings in Planck scale. We solve the RG equations numerically and determine the RG evolution of the couplings of the models. For input parameters, we pick benchmark points for parameters of the model that are  consistent with all constraints considered previously in the paper. Similar analyses with different input parameters have been performed for $U(1)$ extension of SM in Ref\cite{Duch:2015jta}.	The running of couplings up to the Planck scale have been shown in Figures.(\ref{RGE}~.a-c). In Figure.~(\ref{RGE}~.d), we compared running of $\lambda_H$ in the model with SM Higgs coupling. As it is known, SM Higgs coupling will be negative for $\mu> 10^{10}~\rm GeV$. This means, it is not possible to establish all three conditions (perturbativity, vacuum stability and positivity) simultaneously in any scale. It is remarkable that  the SM stability problem (positivity of  $\lambda_H$) is solved in the model. This issue arises that  $\lambda_{SH}$ changes very little in our model. This leads small changes for $\lambda_H$ and as a result, $\lambda_H$ remains positive until the Planck scale.

\begin{figure}
	\begin{center}
		\centerline{\hspace{0cm}\epsfig{figure=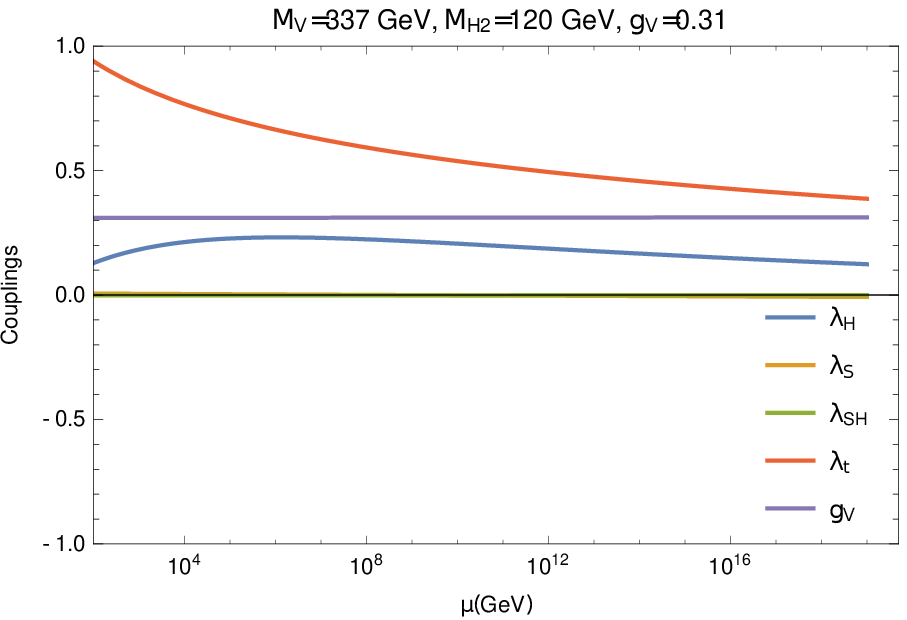,width=8cm}\hspace{0.5cm}\hspace{0cm}\epsfig{figure=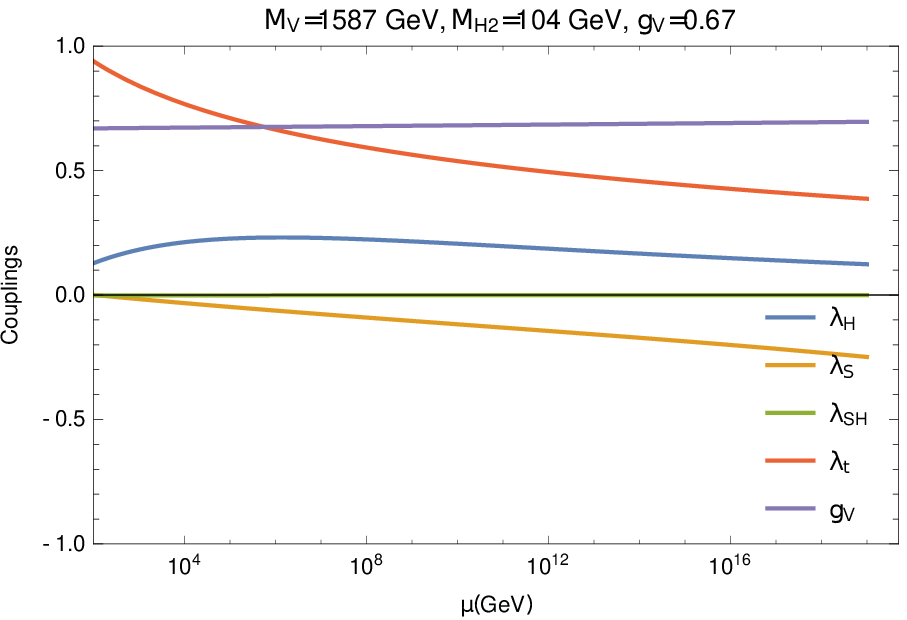,width=8cm}}
		\centerline{\vspace{0.2cm}\hspace{1.5cm}(a)\hspace{8cm}(b)}
		\centerline{\hspace{0cm}\epsfig{figure=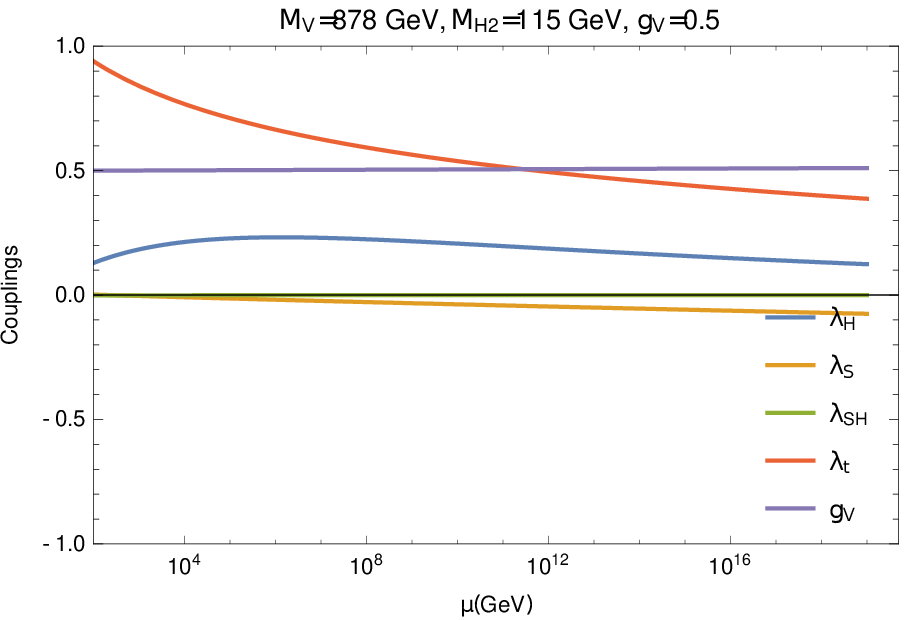,width=8cm}\hspace{0.5cm}\hspace{0cm}\epsfig{figure=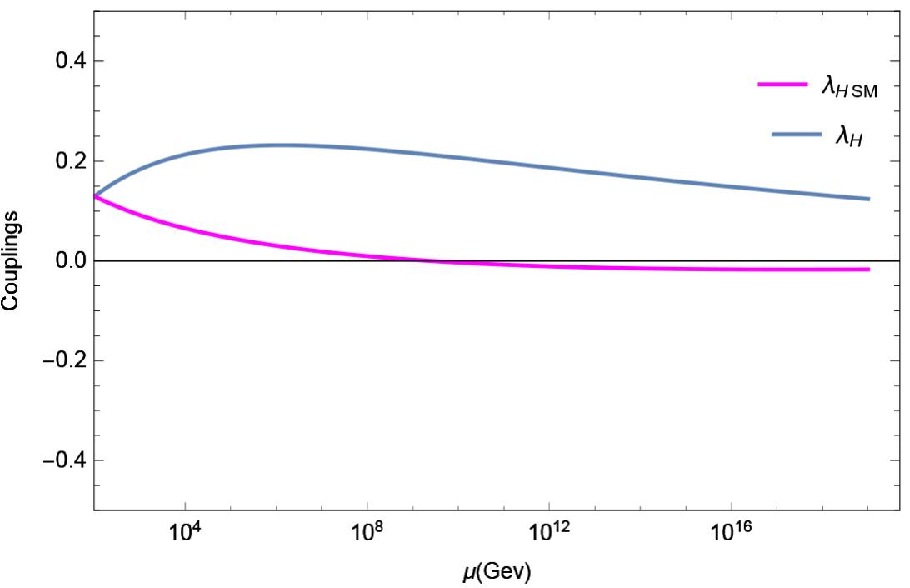,width=8cm}}		
		\centerline{\vspace{0.2cm}\hspace{1.5cm}(c)\hspace{8cm}(d)}
		\centerline{\vspace{-0.2cm}}
		\caption{Running of couplings of the model up to Planck scale. We select sample point in which all the experimental constraints considered in the paper are satisfied.} \label{RGE}
	\end{center}
\end{figure}

\section{Conclusions} \label{sec7}

We proposed a model to explain the W boson mass anomaly reported by the CDF-II collaboration. We studied an $U(1)$ extension of the SM including a VDM candidate and a scalar mediator. In the model, there is no kinetic mixing between the VDM field and SM Z-boson, but scalar field exchange between SM and dark side. To explain the W mass anomaly one needs extra degrees of freedom that affects on W-boson mass. In the model, the one-loop corrections induced by the new
scalar can shift the W boson mass. We have also imposed constraints on the Higgs mixing angle and other parameters of the model by investigating of relic density of DM, invisible Higgs decay mode at LHC and direct detection of DM. We have shown that the model for a large part of VDM mass values, scalar mediator mass range $100~\rm GeV \lesssim M_{H_2}\lesssim 124~\rm GeV$ and $0<g_v\leqslant 1$ can explain W-mass anomaly and satisfy all experimental constraints.

We have also investigated the effects and consequences of the RGE at one-loop order on the model. It was shown that adding of new fields to SM, changes  positivity and stability conditions of SM Higgs as it will be stable up to Planck scale.

\section*{Acknowledgments}
We would like to thank Dr. Ahmad Mohamadnejad for helping with micrOMEGAs code issues.

\bibliography{References}
\bibliographystyle{JHEP}

\end{document}